\documentclass[11pt,twoside]{article}
\usepackage{asp2010}
\resetcounters
\begin{document}
\title{Volume Rendering of AMR Simulations}
\author{Marc~Labadens$^1$, Daniel~Pomar\`{e}de$^1$, Damien~Chapon$^1$, Romain~Teyssier$^1$, Frederic~Bournaud$^1$, Florent~Renaud$^1$, Nicolas~Grandjouan$^2$
\affil{$^1$DSM/IRFU - Centre d'Etudes de Saclay, 91191 Gif-sur-Yvette}
\affil{$^2$LULI - Ecole Polytechnique, Palaiseau}}
\begin{abstract}
High-resolution simulations often rely on the Adaptive Mesh Resolution (AMR) technique to optimize memory consumption versus attainable precision. While this technique allows for dramatic improvements in terms of computing performance, the analysis and visualization of its data outputs remain challenging. The lack of effective volume renderers for the octree-based AMR used by the RAMSES simulation program has led to the development of the solutions presented in this paper. Two custom algorithms are discussed, based on the splatting and the ray-casting techniques.  Their usage is illustrated in the context of the visualization of a high-resolution, 6000-processor simulation of a Milky Way-like galaxy. Performance obtained in terms of memory management and parallelism speedup are presented.
\end{abstract}
\section{Introduction}
\begin{figure}[h!]
	\center
	\includegraphics[width=4cm]{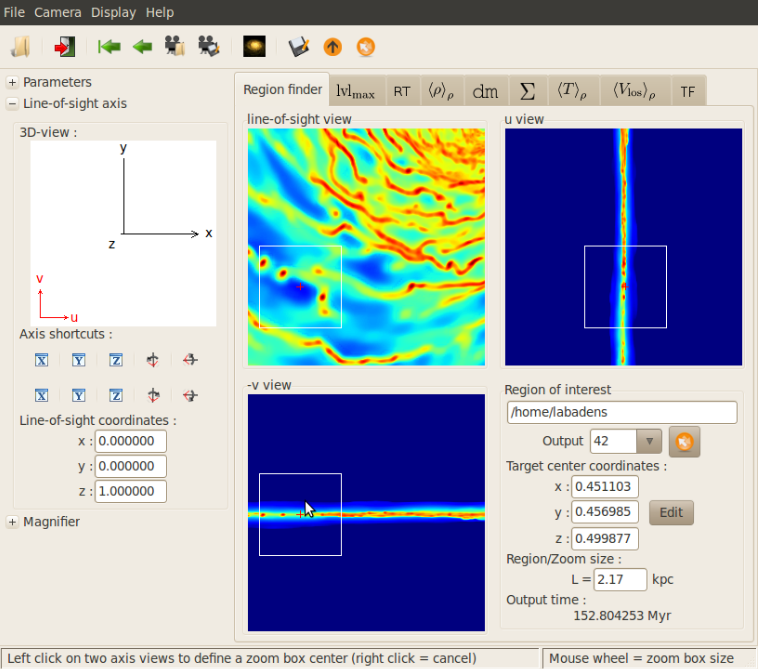}
	\hspace{4mm}
	\includegraphics[width=5cm]{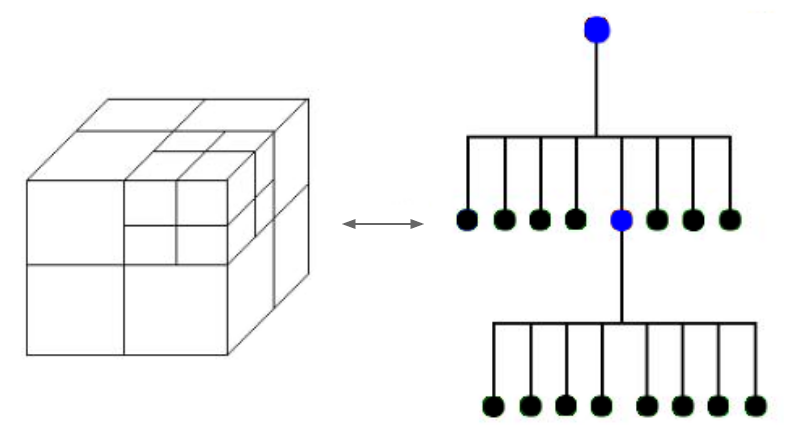}
	\caption{\label{amrviewer} The amrviewer PYMSES GUI (left) and a level two octree illustration (right).}
\end{figure}

Computer simulation is a dynamic area of research in today's astrophysics. In the context of the COAST « Computational Astrophysics » project\footnote{http://irfu.cea.fr/Projets/COAST/}, a wide range of fundamental issues in astrophysics are addressed using the RAMSES simulation program, including the formation of large-scale structures, galaxies, interstellar medium and supernovae. Within this project, the RAMSES code is a grid-based hydrodynamic solver using an octree-based Adaptive Mesh Resolution (AMR) representation \citep{2002A&A...385..337T}. This code is built to be massively parallel with the Message Passing Interface (MPI) library. It involves a fully threaded tree, where every cell has a reference to the neighbors of the cell.
The Milky Way simulation is a great example of RAMSES being used to study galaxy evolution, structure, dynamics and its interstellar medium. Thanks to TGCC/GENCI-2192 and PRACE-0283 allocations, it uses 6080 processors on the new French Petaflop High-Performance Computer (HPC) called "Curie". It involves an octree refined up to level 22, with around 200 million AMR cells. Every cell holds scalars and vectors physical parameters like gas pressure, density and velocity.  This simulation produced around 30 TB of data (around 80 GB of data produced by output, with 400 time outputs). This very large dataset is not convenient for easy visualization purposes as it cannot fit in a fast computer's Random Access Memory (RAM).
The lack of volume rendering software working on the octree AMR data structure leads us to research our own visualization solutions. We currently use a specific code, which is mainly written in the Python programming language. It was named PyMSES\footnote{http://irfu.cea.fr/Projets/PYMSES/intro.html}, an acronym that stands for Python Modules for RAMSES. It makes abundant use of the fast Python numerical library Numpy which is written in C. It also uses Cython and pure C code for fast customized computation on the octree. Cython\footnote{http://cython.org/} is a program that compiles typed Python code into C code to speed up the execution of Python code. This is necessary to boost our volume rendering code performance. A careful analysis of the code was conducted in order to identify its main sections where optimization was needed to improve its overall performance.
In terms of portability, it has to be stressed that this code may be executed on the same platform as the simulation, without any specific hardware GPU resource needed (pure software implementation). Avoiding moving data from the simulation HPC to a specific visualization machine is important as moving such big datasets takes quite a lot of time, computer and network resources. A Graphic User Interface (GUI) has been developed with wxPython, which can be used as an efficient complement to Python scripts. The iterative zooming in process suggested here is a good way to explore an octree AMR in depth(see Figure \ref{amrviewer}).
\section{Volume rendering algorithms.}
The octree data structure is very convenient to adjust the level of resolution of the visualization. Highest level of resolution cells are discarded when these cells are too small compared to the image pixel size. The maximum octree depth level read is therefore defined by the camera. Every cells ranked with this maximum level of resolution are considered as leaves. Doing this, reading every cells of the dataset is never required.
\subsection{The splatting technique}
\begin{figure}[h!]
	\center
	\includegraphics[width=6cm]{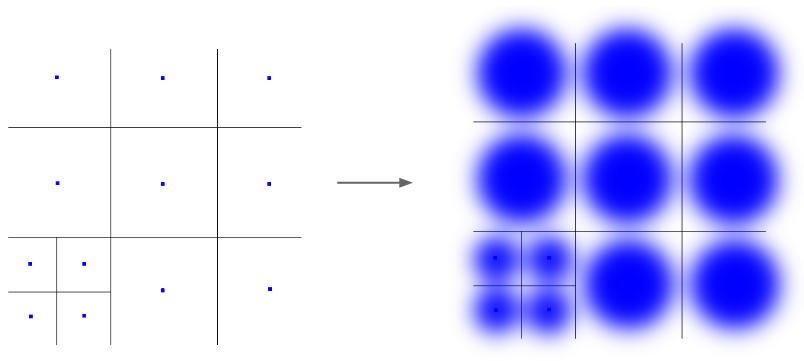}
	\caption{\label{splatting} With the splatting technique every volume element is splatted on the image like a Gaussian disk. With high-resolution images, the result is close to volume ray casting image with a small adaptive Gaussian blur.}
\end{figure}

This volume rendering technique has already been described by \cite{splatting_technique}. AMR cells are converted into points with additional size information depending on the AMR cell size. A different map is computed for every AMR level involved in the camera defined image. For every level, corresponding AMR cell center points are projected on the projection plan. Then, a 2D histogram is computed for every pixel of the image, weighted with the scalar physical value observed. Those 1D projected points are eventually transformed into 2D disk by doing a convolution product with a 2D Gaussian kernel. This step is done using the Fast Fourier Transform (FFT), as a multiplication corresponds to a convolution product in Fourier's space. This was already explained by \cite{fft_splatting}. This implementation currently runs on the CPU using the Numpy's FFT module. Finally all maps obtained for every level are added, which works well as we aim to produce basic summation plot. The interest is that resulting images are close to our ray-traced Maximum Intensity Projection (MIP) maps.
This splatting with FFT transform implementation produces a semi-interactive visualization tool \citep{chapon_phdthesis}.
One advantage is that a complex AMR is easily processed. Figure \ref{splatting} represents the splatted disk sizes that just have to be adjusted with the corresponding volume element size.
On the other hand the grid alignment may appear on resulting images. One possible solution is to increase splatted disk/point sizes. Another solution is to add a random shift to point position, but those quick solutions lead to lower image quality.
\subsection{The volume ray casting technique}
\begin{figure}[ht!]
	\center
	\includegraphics[width=8cm]{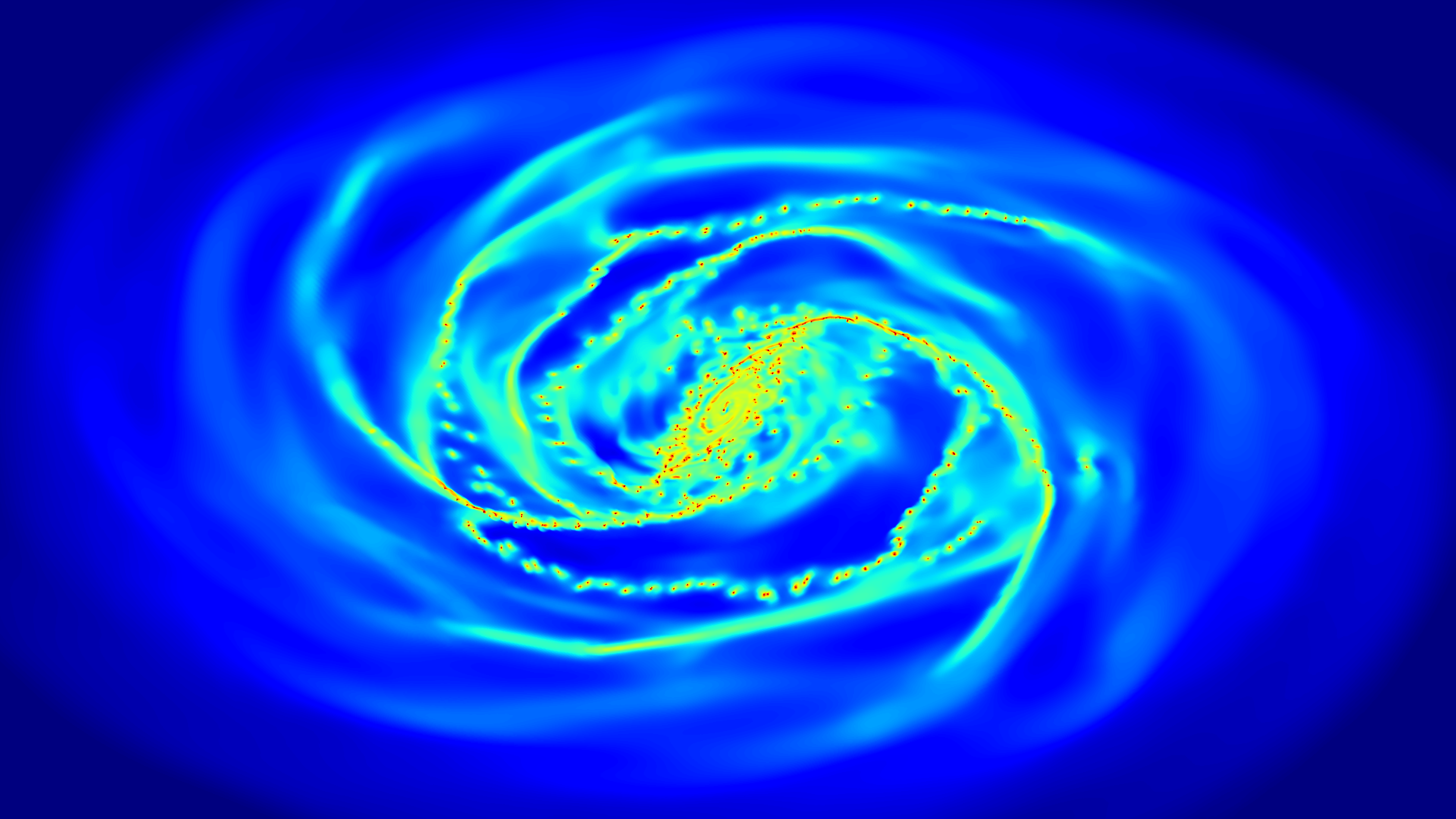}
	\caption{\label{MilkyWay} PyMSES volume ray casting image of a Milky Way simulation output. Every output produces around 80 GB of data distributed among more than 24 thousands files. We visualize here an octree structure refined up to the level 22, with around 200 million of AMR cells.}
\end{figure}

Volume ray casting has been previously explained by \cite{ray_tracing_technique}. For each pixel of the image, one ray is shot through our volume. We now describe our simplest implementation of our adaptive ray tracing. Firstly, the intersections between the grid and the ray are found. Using basic 3D geometry calculation, the coordinates (x,y,z) of the entry and exit points of the ray in the cell faces are computed. The algorithm starts from the root and then goes down the octree. This "top-down" traversal of the tree stops when a leaf cell is found. The cell physical value observed is then multiplied by the length between the entry and exit points into this cell faces. Every contribution along the ray is then added to the resulting pixel value, without taking into account any ordering. This is convenient to process large dataset as this algorithm is embarrassingly parallel.
Compared to the splatting technique, the advantage of direct volume ray casting is that it produces a better image quality (as there are no splatting artifacts, see Figure \ref{MilkyWay}). On the other hand this technique is more computationally intensive, which means our software implementation is not interactive on a large dataset like our Milky Way simulation. When images involving various levels of resolution are computed, some aliasing can be seen on coarse mesh resolution area. Indeed, a low simulation resolution area leads to big pixels. This is why an adaptive Gaussian blur filter can be set as an image post-processing option in our software to avoid this aliasing in coarse mesh regions. The principle is simply to adjust the filter size to local image resolution.
\section{Parallelization}
\subsection{Mpi4py.}
We first tried the Mpi4py library, which regroups Python bindings of the MPI library \citep{mpi4py}. Data is distributed among processing units with a "sort last" strategy, as we want to load those very large datasets from disk in parallel. An advantage is that it can use as many processors as there are available in the cluster. This is useful to make the most of supercomputers with distributed memory. We unfortunately found an increasing initialization time with the number of processors used. Thus there is no interest yet in using too many processors unless there is a requirement for computing lots of independent images, such as for movie creation.
\subsection{Python multiprocessing.}
One advantage of Python multiprocessing is that it is easy to implement and deploy. One drawback is that there is not yet shared memory with the Numpy arrays, which makes load balancing harder as specific arrays are then needed for each process. This code is also limited to one HPC node only if it does not use the MPI library: the number of processors has to be subsequently adjusted. The RAM available is also more limited (compared to using many HPC nodes with distributed RAM), which reduces the possibility of data caching.
In order to get a nice load balancing between processes during our parallel volume ray-casting rendering, we first tried an easy-to-implement strategy. The idea was to start more processes than the number of processors. The Operating System (OS) task scheduling will therefore automatically associate every process to the most available processor. A drawback of this strategy is that, depending on the OS task scheduler performance, we will not get the best results and there is often an OS-defined process number limit, which, for instance, limited us when we tried to run our code with a 64-core computer. Eventually a dynamic load balancing code that creates an adjustable number of worker processes was used. In this case there is a master process that dynamically sends some work to do to available worker processes until the rendering is finished. Again data here is distributed between workers with a "sort last" strategy.

\subsection{Multiprocessing versus Mpi4py}
\begin{figure}[h!]
	\center
	\includegraphics[width=6cm]{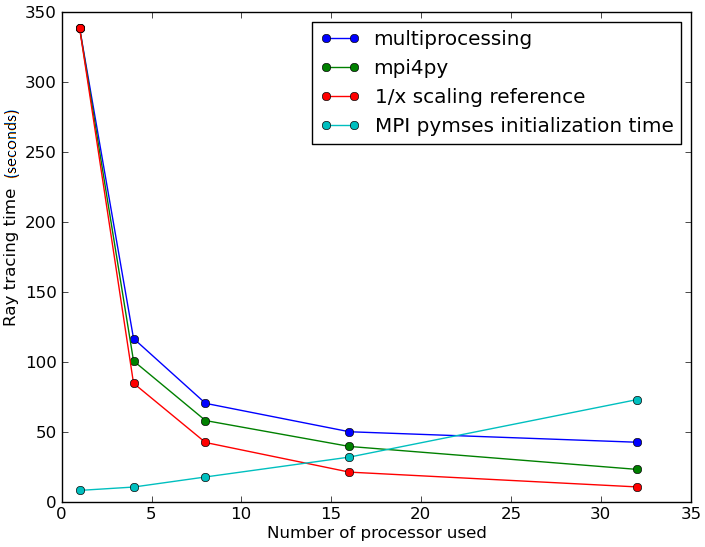}
	\caption{\label{parallel_bench} A full HD PyMSES parallel volume ray casting benchmark on a Milky Way simulation output example. It takes time to read through our large dataset, but getting the code parallel with 16 or 32 processors helps a lot.}
\end{figure}

\begin{figure}[h!]
	\center
	\includegraphics[width=9cm]{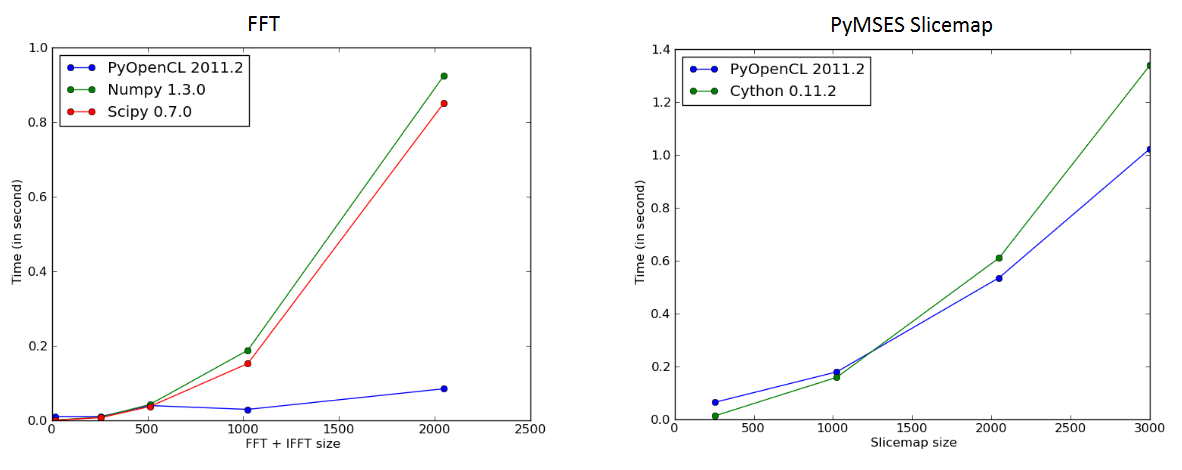}
	\caption{\label{OpenCL} Our OpenCL benchmarks showed some interest only for specific cases with long parallel computation. Loading and moving the right part of the data from disk up to the processing unit is again the main performance bottleneck.}
\end{figure}

The main speed increase comes from parallel I/O and there is not any a big gap between those two possibilities in terms of performance (see Figure \ref{parallel_bench}). The performance bottleneck to visualize our simulation results is indeed still the reading part as large datasets are loaded from hard disks drive. Getting an image from an 80 GB output takes some time and the advantage of having our visualization code in parallel is important. Loading data in parallel is the key point as it is the bottleneck and really it is also better to avoid moving large data too much.
A good point is that HPCs like Curie have good file systems mounted on many hard disks to get some decent I/O performances. For instance, it takes approximately 5 min to load data and process a first single image through our 3D AMR simulation box using the mono thread version of our code. Using an HPC node with 16 processors, we get exactly the same first image in around 20 s, which is much better to get a quick look at a new part of the simulation.
We have also done some PyOpenCL attempts to speed up parallel computations with an Nvidia Quadro FX 1800 graphics card.
The OpenCL FFT available and a custom PyMSES Slicemap implementation (which is a simple tree search algorithm) were tested. It resulted in some speed increase while using the graphics card, but only on some specific cases. Indeed the code has to give the graphics card a lot of work to do to balance the data-to-device transfer overhead (see Figure \ref{OpenCL}).
\section{Visualization with VisIt.}
\begin{figure}[h!]
	\center
	\includegraphics[width=4cm]{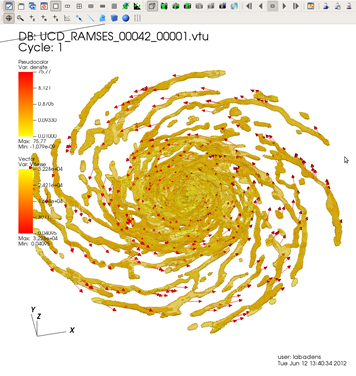}
	\hspace{6mm}
	\includegraphics[width=4cm]{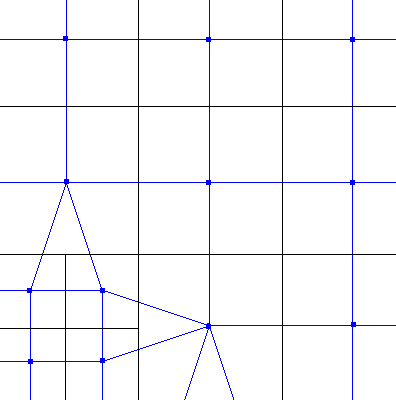}
	\caption{\label{visitFig} Visit Unstructured Cell Data data of a galaxy simulation (on the left) and a two-dimensional quadtree dual mesh illustration (on the right).}
\end{figure}

Another idea to visualize our octree mesh simulation results was to reuse some of the open source advanced visualization software that is currently available, such as VisIt\footnote{https://wci.llnl.gov/codes/visit/}. We had to solve the problem of converting our data into a data format compatible with the visualization software. The best solution that we currently have is to turn our octree AMR into Unstructured Cell Data (UCD), an existing format that allows complex mesh geometry. As we have cell centered physical data values, the idea is to use the dual mesh to produce our UCD. This current implementation works with VisIt (Figure \ref{visitFig}). A nice point is that VisIt can be fully run in parallel with MPI. Advanced VisIt features like filtering, volume rendering, and isosurface reconstruction possibilities can then be reused.
\section{Conclusions and perspectives.}
A specific open source octree geometry parallel visualization software has been developed with efficient Python technologies. It is available online so one can freely use and/or modify the code. Some more work is still needed to make the most of graphics cards. Future work with VisIt involves implementing "in situ" visualization code inside the RAMSES simulation code with the "Libsim" library. The main advantage of this technique is to avoid the current I/O visualization bottleneck. This next step requires the simulation code to be slightly changed to include some smart data transformation and communication with the VisIt software.
\bibliography{author}

\begin{thebibliography}{}
\expandafter\ifx\csname natexlab\endcsname\relax\def\natexlab#1{#1}\fi
\expandafter\ifx\csname url\endcsname\relax
  \def\url#1{\texttt{#1}}\fi
\expandafter\ifx\csname urlprefix\endcsname\relax\def\urlprefix{URL }\fi
\providecommand{\eprint}[2][]{\url{#2}}

\bibitem[{Chapon(2011)}]{chapon_phdthesis}
Chapon, D. 2011, Ph.D. thesis, Universit\'{e} Paris 7

\bibitem[{Dalcin(2012)}]{mpi4py}
Dalcin, L. 2010-2012, Mpi for python

\bibitem[{{Levoy}(1990)}]{ray_tracing_technique}
{Levoy}, M. 1990, IEEE Computer Graphics, 33

\bibitem[{{Teyssier}(2002)}]{2002A&A...385..337T}
{Teyssier}, R. 2002, \aap, 385, 337. \eprint{arXiv:astro-ph/0111367}

\bibitem[{Thomas~Jansen(2004)}]{fft_splatting}
Thomas~Jansen, N. H. E.~K., Bartosz von Rymon-Lipinski 2004, VMV

\bibitem[{{Westover}(1990)}]{splatting_technique}
{Westover}, L. 1990, Computer Graphics, 24, 367

\end{thebibliography}
\end{document}